\pdfoutput=1
\documentclass[twocolumn,superscriptaddress,aps,preprintnumbers,amsmath,amssymb,prd,nofootinbib]{revtex4-2}

\usepackage{amsmath}
\usepackage{amssymb}
\usepackage{amsfonts}
\usepackage{graphicx}
\usepackage{xcolor}
\usepackage{xfrac}
\usepackage{comment}
\usepackage{pifont}
\usepackage{physics}
\usepackage{fourier}
\usepackage{hyperref}
\usepackage{bm}
\usepackage{enumitem}

\definecolor{rossoferrari}{HTML}{D9073D}
\definecolor{mediumblue}{HTML}{0000CD}
\definecolor{forestgreen}{HTML}{228B22}
\definecolor{desy_blue}{HTML}{009EE2}
\definecolor{desy_orange}{HTML}{FD8800}
\definecolor{light_pink}{rgb}{1,0.4,0.4}
\definecolor{light_blue}{rgb}{0.284602,0.317763,0.963947}
\hypersetup{setpagesize=false,bookmarksnumbered=true,bookmarksopen=true, colorlinks=true,linkcolor=light_blue,urlcolor=rossoferrari,citecolor=rossoferrari,linktocpage=false}

\usepackage{slashed}

\renewcommand{\thefootnote}{\fnsymbol{footnote}}

\newcommand{\bea}{\begin{array}}
\newcommand{\eea}{\end{array}}
\newcommand{\beq}{\begin{eqnarray}}
\newcommand{\eeq}{\end{eqnarray}}

\newcommand{\lmk}{\left(}  
\newcommand{\rmk}{\right)}
\newcommand{\lkk}{\left[}  
\newcommand{\rkk}{\right]}

\newcommand{\Mpl}{M_{\rm Pl}}

\def\eq#1{Eq.~(\ref{#1})}

\definecolor{orange}{RGB}{255,100,0}
\definecolor{rosepink}{RGB}{248,100,100}

\begin{document}

\title{
Maximal GW amplitude from bubble collisions in supercooled phase transitions
}

\author{Masaki Yamada}
\email{m.yamada@tohoku.ac.jp}
\affiliation{Department of Physics, Tohoku University, Sendai, Miyagi 980-8578, Japan}

\preprint{TU-1275}

\date{\today}


\begin{abstract}
\noindent
We extend analytic formulas for the gravitational-wave (GW) spectrum from first-order phase transitions to include cosmic expansion under the thin-wall and envelope approximations. We demonstrate that even for strongly supercooled transitions the GW amplitude is bounded from above. This conclusion is explicitly verified for several representative nucleation histories, including delta-function, power-law, and power-exponential types. Moreover, the spectral shape, amplitude, and peak frequency remain largely unaffected by the details of the nucleation rate once expressed in terms of the conformal variables evaluated at an appropriately defined characteristic collision time.
\end{abstract}

\maketitle

\renewcommand{\thefootnote}{\arabic{footnote}}
\setcounter{footnote}{0}

\section{Introduction}

Gravitational-wave (GW) astronomy has entered the precision era with detections of binary black holes~\cite{LIGOScientific:2016aoc,LIGOScientific:2016sjg,LIGOScientific:2017vwq}. Pulsar timing arrays have also reported a nanohertz GW background~\cite{EPTA:2023fyk,NANOGrav:2023gor,Reardon:2023gzh,Xu:2023wog}, and upcoming space- and ground-based interferometers will probe an even broader frequency range~\cite{Janssen:2014dka,LISA:2017pwj,Kawamura:2011zz,Kawamura:2020pcg,Harry:2006fi,Punturo:2010zz,Maggiore:2019uih,Reitze:2019iox,Somiya:2011np,KAGRA:2020cvd,Colpi:2024xhw}. 
This progress provides opportunities to explore fundamental physics through GW observations, since cosmological sources-such as inflation~\cite{Starobinsky:1979ty}, topological defects~\cite{Vilenkin:1981bx,Vachaspati:1984gt}, and first-order phase transitions (FOPTs)~\cite{Witten:1984rs}-can produce distinctive signals linked to physics beyond the Standard Model.

FOPTs arise generically in simple extensions of the Standard Model and in many dark-sector models~\cite{Quiros:1999jp,Morrissey:2012db,Schwaller:2015tja,Athron:2023xlk}. The predicted spectrum depends on bubble-wall dynamics and plasma interactions: for non-runaway walls, long-lived sound waves dominate and turbulence can contribute~\cite{Hindmarsh:2013xza,Hindmarsh:2017gnf,Hindmarsh:2019phv,Caprini:2009yp,RoperPol:2019wvy}; for (nearly) runaway walls, bubble collisions are the leading source~\cite{Kosowsky:1992rz,Kosowsky:1992vn,Kamionkowski:1993fg,Huber:2008hg,Bodeker:2009qy,Weir:2016tov,Jinno:2016vai,Konstandin:2017sat,Jinno:2017fby,Cutting:2018tjt,Megevand:2021juo,Megevand:2021llq,Cai:2023guc}. In particular, Jinno and Takimoto derived analytic expressions for energy-momentum correlators under thin-wall and envelope approximations~\cite{Jinno:2016vai} (see also~\cite{Jinno:2017fby}), yielding multi-dimensional integrals that are numerically tractable and capture universal spectral features.

This Letter extends the analytic formalism of bubble collisions for runnaway walls to an expanding FLRW background under the thin-wall and envelope approximations, while consistently accounting for the Hubble expansion throughout the transition.%
\footnote{Related work~\cite{Zhong:2021hgo} found that the Minkowski-spacetime assumptions can overestimate amplitudes for small $\beta/H$. Our treatment clarifies the timing implicit in $\alpha$ and accommodates general nucleation profiles.}\footnote{See also recent studies of gravitational backreaction of large vacuum bubbles~\cite{Jinno:2024nwb,Giombi:2025tkv}, which analyze fluid-profile distortions relevant for finite-width sound shells.}
The extension is crucial for strongly supercooled FOPTs with durations comparable to $H^{-1}$. We show that cosmic expansion redshifts the sourced tensor modes and demonstrate an upper bound on the GW amplitude from bubble collisions, sharpening earlier qualitative arguments~\cite{Ellis:2018mja}. 
As concrete applications, we examine delta-function, power-law, and power-exponential forms of the bubble nucleation rate and evaluate the corresponding GW spectra.
We find that, after factorizing parameter dependence by the inverse duration $\beta^{-1}$ and the Hubble parameter evaluated at an appropriately defined characteristic collision time, the spectral shape, amplitude, and peak frequency become largely insensitive to details of the nucleation rate. These results provide a qualitative estimate of prospective signals from supercooled transitions~\cite{Ellis:2019oqb,Lewicki:2019gmv,Lewicki:2020jiv,Ellis:2020nnr,Lewicki:2022pdb} (see also~\cite{Konstandin:2011dr,Megevand:2016lpr,Ellis:2018mja,Hashino:2018wee,Brdar:2018num,Fujikura:2019oyi}) with $\mathcal{O}(1)$ accuracy.

\section{Characteristic parameters for bubble nucleation}
\label{sec:expansion}

We consider a false vacuum at high temperature that undergoes a FOPT as the temperature drops below a critical value. True-vacuum bubbles nucleate, expand, and collide until percolation; the resulting collisions source GWs. Throughout this letter, we take bubble walls to expand at (nearly) the speed of light.

We work in an FLRW background with scale factor $a(\tau)$ and conformal time $\tau$. The bubble nucleation rate per unit \emph{comoving} volume and \emph{conformal} time is defined by 
\begin{align}
 \tilde{\Gamma}(\tau) \equiv a^4 (\tau) \Gamma(\tau) \,,
 \label{eq:gammatilde}
\end{align}
where $\Gamma$ is the nucleation rate per unit physical volume and physical time.

The false-vacuum survival probability is $P_1(\tau)=e^{-I_1(\tau)}$, with
\begin{align}
 &I_1(\tau) 
 = \int_0^{\tau} d\tau' 
 \frac{4\pi}{3} \lmk \tau - \tau' \rmk^3 \tilde{\Gamma} (\tau') \,.
 \label{eq:I}
\end{align}
For notational simplicity we take nucleation in $\tau\in(0,\infty)$. We define the characteristic collision time $\tau_*$ by $I_1(\tau_*)=1$, and denote the Hubble and conformal Hubble parameters at $\tau_*$ by $H_*$ and $\tilde H_*\equiv a(\tau_*)H_*$, respectively.

The (conformal) growth rate of the true-vacuum fraction at $\tau_*$ is defined by 
\begin{align}
\tilde{\beta} \equiv - \frac{d \ln P_1(\tau_*)}{d\tau} = \frac{d I_1(\tau_*)}{d \tau} \,.
\label{eq:beta}
\end{align}
We also define $\beta \equiv \tilde{\beta}/a(\tau_*)$, the inverse physical duration of the transition. The ratio $\beta/H_*=\tilde{\beta}/\tilde{H}_*$ measures the importance of expansion: Hubble effects are negligible for $\beta/H_*\gg 1$, while here we allow $\beta/H_* = \mathcal{O}(1)$.

We neglect metric backreaction and approximate the background as homogeneous and isotropic FLRW. We mainly consider cases where the dominant component of the Universe (e.g., the visible sector) is sequestered from the nucleating sector: a dark sector undergoes the FOPT while remaining decoupled during the transition. This permits thin-wall bubbles with large latent heat without significant backreaction on expansion. 

Let $\rho_0(\tau)$ denote the vacuum energy difference (or, more precisely, the potential energy at the tunneling point), which may be time dependent. Denoting the energy density for background components in the Universe as $\rho_{\rm tot}(\tau)$ ($= 3 H^2(\tau) \Mpl^2$), 
we define
\begin{align}
\alpha_*
&\equiv \frac{\rho_0 (\tau_*)}{\rho_{\rm tot}(\tau_*)}\,.
\label{eq:alpha}
\end{align}
The backreaction is negligible for $\alpha_* \ll 1$, which we assume throughout this Letter.%
\footnote{If $\alpha_*\gtrsim 0.01$ at $\tilde{\beta}/\tilde{H}_*\simeq3$, some regions remain in the false vacuum and undergo eternal inflation, appearing as black holes to external observers~\cite{Jinno:2023vnr}.}
The efficiency factor $\kappa(\tau)$ is the fraction of vacuum energy deposited into the bubble wall and may be time dependent. We denote $\kappa_* \equiv \kappa(\tau_*)$.

With these parameters, the stochastic GW spectrum reads
\begin{align}
\Omega_{\rm GW}(\tau,k) 
&\equiv \frac{1}{\rho_{\rm tot}(\tau)} \frac{d\rho_{\rm GW}(\tau,k)}{d\ln k} \nonumber \\
&= \kappa_*^2 \alpha_*^2 \left(\frac{\tilde{H}_*}{\tilde{\beta}}\right)^2 \lmk \frac{a^4 (\tau_*) \rho_{\rm tot}(\tau_*)}{a^4 (\tau) \rho_{\rm tot}(\tau)} \rmk
\Delta(k, \tilde{\beta}) \,,
\label{eq:Omega_Delta}
\end{align}
with comoving wavenumber $k$. The function $\Delta$ is given below. 
Note that $\beta/H_*=\tilde{\beta}/\tilde{H}_*$.

For reference, for a transition during radiation domination the present frequency and amplitude redshift factor are
\begin{align}
f
&\simeq 1.65 \times 10^{-5} \, {\rm Hz}
\left( \frac{k}{2 \pi \tilde{\beta}} \right) \left( \frac{\tilde{\beta}}{\tilde{H}_*} \right)
\left( \frac{T_*}{10^2 \, {\rm GeV}} \right)
\left( \frac{g_*}{100} \right)^{\frac{1}{2}}
\left( \frac{g_{*s}}{100} \right)^{- \frac{1}{3}} \,,
\label{eq:f_present}
\end{align}
and
\begin{align}
\lmk \frac{a^4 (\tau_*) \rho_{\rm tot}(\tau_*)}{a^4 (\tau) \rho_{\rm tot}(\tau)} \rmk &\simeq  1.67\times 10^{-5} 
\left( \frac{g_*}{100} \right)
\left( \frac{g_{*s}}{100} \right)^{-\frac{4}{3}} \,,
\label{eq:Omega_present}
\end{align}
where $T_*$ is the temperature at $\tau=\tau_*$, and $g_*$ and $g_{*s}$ are the effective relativistic degrees of freedom for energy density and entropy. If GWs are produced before the completion of reheating, additional dilution and redshift factors should be included in Eqs.~(\ref{eq:f_present}) and (\ref{eq:Omega_present}). However, we note that \eq{eq:Omega_Delta} and our analytic formula in Sec.~\ref{sec:analytic} apply to general expansion histories.

\section{Analytic formula for GW spectrum in an expanding Universe}
\label{sec:analytic}

We compute correlators of the energy-momentum tensor for thin-wall bubbles under the envelope approximation and derive an analytic GW spectrum in an expanding background. The calculation parallels Refs.~\cite{Jinno:2016vai,Jinno:2017fby}, but is formulated in conformal variables and includes explicit redshift factors. Cosmic expansion enters through the changing false-vacuum volume, the vacuum energy, and the redshifting of wall and GW energies/frequencies. A detailed derivation is given in a companion paper~\cite{Yamada:2025cfr}; here we summarize the final expressions.

There are two additive contributions, single-bubble and double-bubble, with $\Delta=\Delta^{(s)}+\Delta^{(d)}$. 
The single-bubble contribution $\Delta^{(s)}$ arises from the correlation of the energy-momentum tensor evaluated at two spacetime points that belong to the same bubble.
This contribution is nonzero because, after collisions with other bubbles, even an individual bubble no longer retains perfect spherical symmetry.
It is given by 
\begin{align}
&\Delta^{(s)} 
(k, \tilde{\beta}) \nonumber\\
&= 6 \tilde{\beta}^2 k^3
\int_{0}^\infty d{\mathcal T}
\int_{0}^{2 \mathcal T} d\tau_d \; 
\lmk \frac{a ({\mathcal T} + \tau_d/2) a({\mathcal T} - \tau_d/2)}{a^2(\tau_*)} \rmk^3 \nonumber\\
&~~~~\times
\cos(k \tau_d) 
 \int_{\tau_d}^{2 {\mathcal T}} d r \, r P_2 ({\mathcal T}, \tau_d, r) 
\int_{0}^{\tau_{xy}} d\tau_n~\tilde{\Gamma}(\tau_n) 
\nonumber\\
&~~~~\times
\frac{l_B^2 \rho_B ({\mathcal T} + \tau_d/2) \rho_B({\mathcal T} - \tau_d/2)}{\kappa^2(\tau_*) \rho_0^2 (\tau_*) r_x r_y }
\nonumber\\
&~~~~\times
\left[
j_0 (k r) {\mathcal S}_0 + \frac{j_1 (k r)}{k r} {\mathcal S}_1 + \frac{j_2 (k r)}{k^2 r^2} {\mathcal S}_2
\right] \,,
\label{eq:Deltas}
\end{align}
where $j_i(x)$ are the spherical Bessel functions, 
\begin{align}
{\mathcal S}_0
&=
r_x^2 r_y^2s_{x\times}^2s_{y\times}^2, 
\label{eq:F0pp}
\\
{\mathcal S}_1
&= 
r_x r_y \left[ 
4c_{x\times}c_{y\times}(r_x^2s_{x\times}^2 + r_y^2s_{y\times}^2)
- 2r_xr_ys_{x\times}^2s_{y\times}^2 
\right] \,,
\label{eq:F1pp}
\\
{\mathcal S}_2
&=
r_x r_y
\bigl[
r_xr_y(19c_{x\times}^2 c_{y\times}^2 - 7(c_{x\times}^2 + c_{y\times}^2) + 3) 
\nonumber\\
&~~~~~~~~~~~~~~~~~~~~~~~~~~~~~~~~~~
- 8c_{x\times}c_{y\times}(r_x^2s_{x\times}^2 + r_y^2s_{y\times}^2)
\bigr] \,,
\label{eq:F2pp}
\end{align}
and
\begin{align}
&r_x
\equiv 
{\mathcal T} + \tau_d/2 - \tau_n,
\quad \quad 
r_y
\equiv 
{\mathcal T} - \tau_d/2 - \tau_n,
\\
&\tau_{xy}
\equiv {\mathcal T} - \frac{r}{2} ,
\\
&c_{x \times}
\equiv
- \frac{r^2 + r_x^2 - r_y^2}{2 r r_x},
\quad \quad 
c_{y \times}
\equiv
\frac{r^2 + r_y^2 - r_x^2}{2 r r_y} \,,
\label{eq:cos}
\end{align}
with $s_{x \times} \equiv \sqrt{1-c_{x \times}^2}$ 
and $s_{y \times} \equiv \sqrt{1-c_{y \times}^2}$. 
This result is obtained from the Fourier transform of the correlation function of the transverse-traceless part of the energy-momentum tensor evaluated at two separated spacetime points. The spatial integrations are carried out except for the time variables and the spatial separation $r$. Since GWs are sourced by correlations of the energy-momentum tensor $T_{ij}$, the integrand contains two factors of the bubble energy density $\rho_B$. 
The factors $P_2$ (explained below) and $\tilde{\Gamma}$ arise from the requirement that two spacetime points lie on the world volume of a single bubble wall, under the condition that they have not yet transitioned to the true vacuum.
The quantities ${\mathcal S}_i$ ($i=0,1,2$) encode geometric information about the allowed regions for bubble nucleation. Finally, the spherical Bessel functions originate from the Fourier transform associated with the computation of the GW spectrum.

The probability $P_2({\mathcal T}, \tau_d,r) = e^{-I_2({\mathcal T}, \tau_d,r) }$ encodes the nucleation history, where 
\begin{align}
&I_2({\mathcal T}, \tau_d,r)
\nonumber\\
&=
 \int_{0}^{\tau_{xy}} d\tau \; \frac{\pi}{3} \tilde{\Gamma}(\tau) 
 \lkk 
 \frac{\lmk r + 2 ({\mathcal T} - \tau) \rmk^2}{4 r} 
 \lmk 3 \tau_d^2 - r^2 + 4 r ({\mathcal T} - \tau) \rmk 
 \rkk
 \nonumber\\ 
 &\;\;\;\;\;\;\;\;\;\;
 + \int_{\tau_{xy}}^{\tau_x} d\tau \; \frac{4\pi}{3} \tilde{\Gamma}(\tau) \lmk {\mathcal T} + \tau_d/2 -\tau \rmk^3  
 \nonumber\\ 
 &\;\;\;\;\;\;\;\;\;\;
+ \int_{\tau_{xy}}^{\tau_y} d\tau \; \frac{4\pi}{3} \tilde{\Gamma}(\tau) \lmk {\mathcal T} - \tau_d/2 -\tau \rmk^3 \,.
 \label{eq:I2xy}
\end{align}
Also, 
\begin{align}
 l_B \rho_B(\tau) &\equiv \frac{1}{4\pi a^3(\tau) (\tau - \tau_n)^2 }
  \nonumber\\ 
 &\;\;\;\;\;\;\;\;\;\;
 \times
\int_{\tau_n}^\tau d \tau'~4 \pi a^3(\tau') (\tau' - \tau_n)^2 \kappa(\tau') \rho_0(\tau') \,,
\end{align}
captures the time dependence of the vacuum-energy difference $\rho_0(\tau)$. The comoving wall width $l_B$ cancels in the final result, allowing the $l_B\to0$ limit.

The double-bubble contribution $\Delta^{(d)}$ arises from the correlation of the energy-momentum tensor evaluated at two spacetime points that belong to different bubbles.
It is given by 
\begin{align}
&\Delta^{(d)}(k, \tilde{\beta})
\nonumber\\
&=
\frac{24 \tilde{\beta}^2 k^3}{\pi}
\int_0^\infty d{\mathcal T}
\int_0^{2 {\mathcal T}} d\tau_d
\lmk \frac{a ( {\mathcal T} + \tau_d/2 ) a({\mathcal T} - \tau_d/2)}{a^2(\tau_*)} \rmk^3
\nonumber\\
&~~~~
\times
\cos(k \tau_d) 
\int_{\tau_d}^{2 {\mathcal T}} dr~r^2 P_2 ({\mathcal T}, \tau_d, r) 
\nonumber\\
&~~~~
\times
\frac{{\mathcal D}^{(d)}_x({\mathcal T},\tau_d, r)
{\mathcal D}^{(d)}_y({\mathcal T},\tau_d, r)}{\kappa^2(\tau_*) \rho_0^2(\tau_*)} \frac{j_2 (k r)}{k^2 r^2} \,,
\label{eq:Deltad}
\end{align}
where 
\begin{align}
&{\mathcal D}^{(d)}_x ({\mathcal T},\tau_d, r)
\nonumber\\
&\equiv
\int_{0}^{\tau_{xy}} d\tau_n~\tilde{\Gamma} (\tau_n) \, \pi r_x^2 l_B \rho_B({\mathcal T} + \tau_d/2)
\lmk c_{x\times}^3 - c_{x\times} \rmk \,,
\\
&{\mathcal D}^{(d)}_y ({\mathcal T},\tau_d, r)
\nonumber\\
&\equiv
\int_{0}^{\tau_{xy}} d\tau_n~\tilde{\Gamma} (\tau_n) \, \pi r_y^2 l_B \rho_B({\mathcal T} - \tau_d/2)
\lmk c_{y\times}  - c_{y\times}^3 \rmk \,.
\end{align}
Since this contribution involves two distinct bubbles, it contains two factors associated with the bubble nucleation rate $\tilde{\Gamma}$. The remaining terms have a qualitative origin similar to those appearing in the single-bubble contribution.

Equations (\ref{eq:Deltas}) and (\ref{eq:Deltad}) give the GW spectrum in an expanding Universe for arbitrary $\rho_0(\tau)$ and $a(\tau)$.

\section{Numerical results}
\label{sec:numerical}

We numerically evaluate Eqs.~(\ref{eq:Deltas}) and (\ref{eq:Deltad}) in concrete examples, using the Vegas algorithm~\cite{Lepage:1977sw}, as implemented in the CUBA library~\cite{Hahn:2004fe}.
Vegas is a Monte Carlo algorithm that employs importance sampling to reduce variance.
All calculations were performed with a requested relative accuracy of $3 \times 10^{-3}$.

We assume radiation domination with $a(\tau)/a_*=\tau/\tau_*$ and take $\kappa(\tau)\rho_0(\tau)=\kappa\rho_0=\text{const.}$
In this case,
\begin{align}
 l_B \rho_B(\tau) = \kappa \rho_0 (\tau-\tau_n) 
 \lkk \frac{1}{6} + \frac{1}{10} \lmk \frac{\tau_n}{\tau} \rmk + \frac{1}{20} \lmk \frac{\tau_n}{\tau}\rmk^2 + \frac{1}{60} \lmk \frac{\tau_n}{\tau} \rmk^3 \rkk \,.
 \nonumber\\
 \label{rhoBex}
\end{align}
The large parentheses on the right-hand side take values $(1/6,1/3)$ for $\tau\in(\tau_n,\infty)$.

\subsection{Case with delta-function nucleation rate}
\label{sec:result_delta}

First, we consider a delta-function nucleation rate:
\begin{align}
 \tilde{\Gamma}(\tau) = n_{\rm nuc} \, \delta(\tau - \tau_{\rm nuc}) \,,
\end{align}
where $n_{\rm nuc}$ is the comoving number density of bubbles and $\tau_{\rm nuc}$ is the nucleation time.%
\footnote{
See Refs.~\cite{Jinno:2023vnr,Zhong:2025xwm} for concrete models of this type of transition. 
This setup can also be regarded as a limiting case of a Gaussian nucleation rate~\cite{Megevand:2016lpr,Jinno:2017ixd,Megevand:2017vtb}.
}
Then
\begin{align}
&\tau_* = \tau_{\rm nuc} + \lmk \frac{3}{4 \pi n_{\rm nuc} } \rmk^{1/3} = \tau_{\rm nuc} + \frac{3}{\tilde{\beta}} \,,
\label{eq:delta_taustar}
\\
&\tilde{\beta} = \lmk 36 \pi n_{\rm nuc} \rmk^{1/3}  \,.
\label{eq:delta_betatilde}
\end{align}
Under radiation domination this implies $\tilde{\beta}/\tilde{H}_* = 3 + \tilde{\beta}\tau_{\rm nuc}$, using $\tilde{H}_*=\tau_*^{-1}$.

The theory has two inputs, $n_{\rm nuc}$ and $\tau_{\rm nuc}$; hence the spectrum depends on $(k,n_{\rm nuc},\tau_{\rm nuc})$. Using Eqs.~(\ref{eq:delta_taustar})-(\ref{eq:delta_betatilde}), one can trade these for $(k, \tilde{\beta},\tau_*)$. 
Since $\Delta^{(s)}$ and $\Delta^{(d)}$ are dimensionless quantities, they can depend only on combinations of dimensionless parameters, such as $k/\tilde{\beta}$ and $\tilde{\beta}\tau_*$ (or equivalently $\tilde{\beta}\tau_{\rm nuc}$).

\begin{figure}
\centering
\includegraphics[width=0.95\hsize]{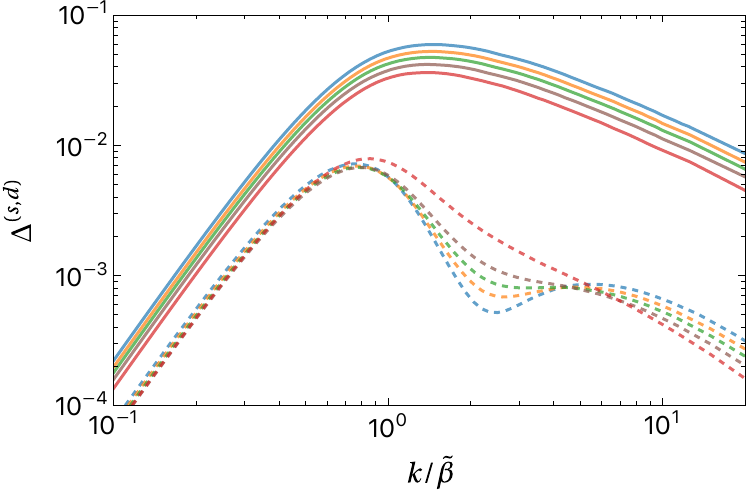}
\caption{
GW spectra $\Delta^{(s)}$ (solid curves) and $\Delta^{(d)}$ (dashed curves) as functions of $k/\tilde{\beta}$ for $\tilde{\beta} \tau_{\rm nuc} = 0.1$ (blue), $0.5$ (orange), $1$ (green), $2$ (brown), and $10$ (red), in the case of delta-function nucleation rate. 
}
\label{fig:delta1}
\end{figure}

Figure~\ref{fig:delta1} shows $\Delta^{(s)}$ (solid) and $\Delta^{(d)}$ (dashed) versus $k/\tilde{\beta}$ for $\tilde{\beta}\tau_{\rm nuc}=0.1$ (blue), $0.5$ (orange), $1$ (green), $2$ (brown), and $10$ (red).
As in the Minkowski-spacetime case, $\Delta^{(d)}$ is subdominant for the whole range of wavenumber.

\begin{figure}
 \centering
 \includegraphics[width=0.95\hsize]{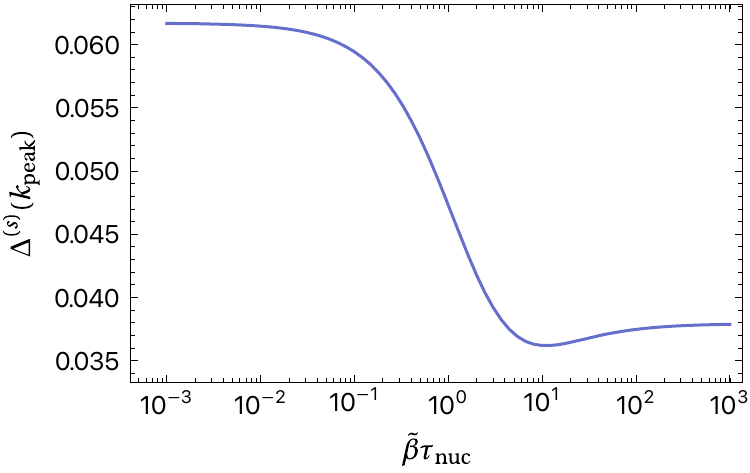}
 \\
 \vspace{0.3cm}
 \includegraphics[width=0.95\hsize]{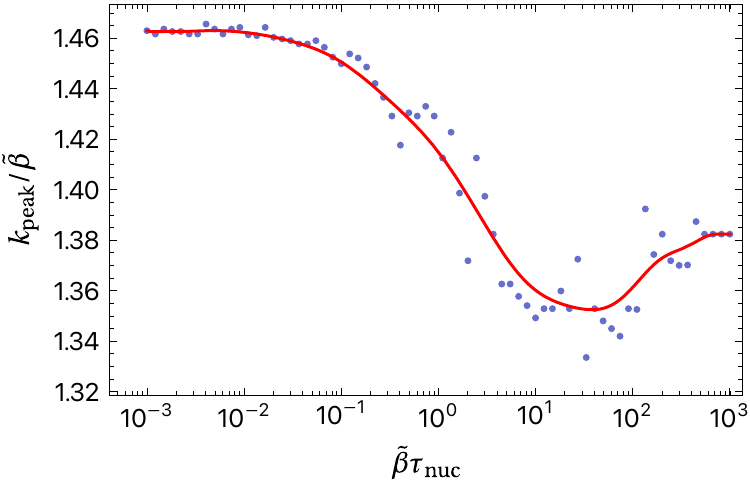}
 \caption{
 Peak amplitude $\Delta^{(s)} (k_{\rm peak},\tilde{\beta})$ (top panel) and wavenumber $k_{\rm peak}/\tilde{\beta}$ (bottom panel) as a function of $\tilde{\beta} \tau_{\rm nuc}$ for the case of a delta-function nucleation rate. 
}
 \label{fig:delta2}
\end{figure}

The peak amplitude $\Delta^{(s)}(k_{\rm peak},\tilde{\beta})$ and position $k_{\rm peak}/\tilde{\beta}$ are shown in Fig.~\ref{fig:delta2} as functions of $\tilde{\beta}\tau_{\rm nuc}\in(10^{-3},10^{3})$. Both approach constants at large and small $\tilde{\beta}\tau_{\rm nuc}$. Because it is difficult to determine $k_{\rm peak}$ precisely in numerical simulations, $k_{\rm peak}/\tilde{\beta}$ scatters by a few percent. The red curve in the lower panel is just a guide to the eye.
In the limit of $\tilde{\beta}\tau_{\rm nuc}\ll1$ (i.e., $\tilde{\beta}/\tilde{H}_*\approx3$),
\begin{align}
 &\Delta^{(s)}(k_{\rm peak}, \tilde{\beta}) \simeq 0.062 \,,
 \label{eq:deltapeak}
 \\
 &\frac{k_{\rm peak}}{\tilde{\beta}} \simeq 1.46 \,,
 \label{eq:kpeak}
\end{align}
corresponding to a super-slow phase transition. 
Additionally, the double-bubble contribution $\Delta^{(d)}$ accounts for a few percent to the total GW spectrum near the peak, $k = k_{\rm peak}$.

From Fig.~\ref{fig:delta1}, the single-bubble spectral shape is nearly universal; changes in $\tilde{\beta}\tau_{\rm nuc}$ mainly rescale the amplitude. 
It exhibits a peak at a characteristic wavenumber $k \sim \tilde{\beta}$, scaling as $k^3$ in the low-$k$ regime and as $k^{-1}$ at high $k$.
These asymptotic behaviors are derived analytically in a companion paper~\cite{Yamada:2025cfr} and are shown to hold for a general class of nucleation-rate functions.

Moreover, from Fig.~\ref{fig:delta2}, the rescale fractor is at most $\mathcal{O}(1)$. 
This follows from our formulas: the $\delta$-function in the integral over $\tau_n$ enforces $\tau_{\rm nuc}\le\tau_{xy}$, i.e., $r\le2(\mathcal{T}-\tau_{\rm nuc})$, and with $0\le\tau_d\le r$ we have $\tau_d,r=\mathcal{O}(\mathcal{T}-\tau_{\rm nuc})$. Omitting $P_2$ would favor large $\mathcal{T}$, but $P_2$ suppresses contributions for $\mathcal{T}-\tau_{\rm nuc}\gtrsim n_{\rm nuc}^{-1/3}\sim\tilde{\beta}^{-1}$ by Eq.~(\ref{eq:delta_betatilde}). Thus the integral is dominated by $\mathcal{T}-\tau_{\rm nuc}\sim\tilde{\beta}^{-1}$, giving $\tau_d,r\lesssim\tilde{\beta}^{-1}$. Neglecting the $(\tau_n/\tau)$ terms in \eq{rhoBex}, the integrand for $\Delta^{(s)}$ then depends only on the combination of $(\mathcal{T}-\tau_{\rm nuc})$, so the explicit $\tau_{\rm nuc}$-dependence drops out through the estimation of $\mathcal{T}-\tau_{\rm nuc}\sim\tilde{\beta}^{-1}$ and $\Delta^{(s)}$ is chiefly a function of $k/\tilde{\beta}$. The residual $(\tau_n/\tau)$ terms induce a weak $\tau_{\rm nuc}$ dependence that particularly vanishes for $\tilde{\beta}\tau_{\rm nuc}\ll1$, consistently with our numerical results in Fig.~\ref{fig:delta2}. 

Moreover, the $r$-integral prefers larger $r$ (up to $\tilde{\beta}^{-1}$) but is damped by oscillatory Bessel functions, leading to a peak near $k\sim\tilde{\beta}$. The same qualitative behavior holds for $\Delta^{(d)}$.

We note that $\mathcal{T}-\tau_{\rm nuc} \sim \tilde{\beta}^{-1}$ arises from the fact that most bubbles nucleate within a timescale of $\tilde{\beta}^{-1}$ just before the end of the phase transition (at $\tau = \tau_*$). The typical bubble size is therefore of order $\tilde{\beta}^{-1}$, since the bubble wall velocity is taken to be the speed of light, implying that $r$ is dominated around $\tilde{\beta}^{-1}$. Whenever these qualitative features hold, the above patterns of GW spectrum should persist for other choices of the nucleation function. Indeed, the power-law and exponential nucleation rates fall into this category and exhibit qualitatively similar behavior, as shown below.

\subsection{Case with power-law nucleation rate}
\label{sec:result_pow}

Next, we consider
\begin{align}
 \tilde{\Gamma}(\tau) 
 &= \tilde{\Gamma}_0 \tau^n \,,
 \label{eq:nucleation-rate-pow}
\end{align}
with constants $\tilde{\Gamma}_0$ and $n$ ($> -1$). 
Note that, in the radiation-dominated Universe, $n=4$ corresponds to a constant nucleation rate per unit physical volume and physical time, $\Gamma(\tau) = ({\rm const.})$~\cite{Coleman:1977py,Callan:1977pt}.

In this case, 
\begin{align}
&\tau_* = \lmk \frac{(1+n)(2+n)(3+n)(4+n)}{8 \pi \tilde{\Gamma}_0} \rmk^{1/(4+n)},
\\
&\tilde{\beta}\tau_* = n+4\,.
\end{align}
Therefore, for a fixed $n$, 
the spectrum depends only on $k/\tilde{\beta}$ and contains no additional free parameters. 

\begin{figure}
 \centering
 \includegraphics[width=0.95\hsize]{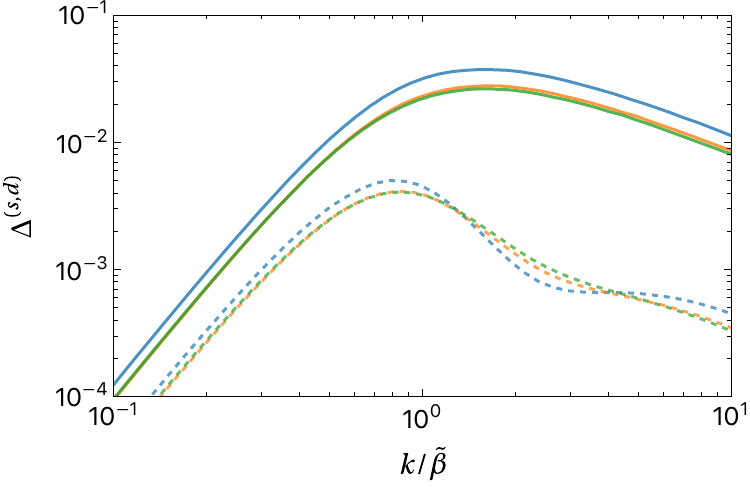}
 \caption{
 Same as Fig.~\ref{fig:delta1} but for a power-law nucleation rate. 
 We take $n =0$ (blue), $2$ (orange), and $4$ (green). 
}
 \label{fig:pow}
\end{figure}

Figure~\ref{fig:pow} shows $\Delta^{(s)}$ (solid) and $\Delta^{(d)}$ (dashed) spectra for $n=0$, $2$, and $4$. 
The overall spectral shape is again very similar across all cases, even for $n=0$ (corresponding to $\tilde{\beta}\tau_* =4$). 
The peak amplitude and frequency of the total spectrum are
\begin{align}
 &\Delta (k_{\rm peak}/\tilde{\beta}) \simeq 0.039 \,,
 \\
 &\frac{k_{\rm peak}}{\tilde{\beta}} \simeq 1.44 \,.
\end{align}
for $n=0$ and 
\begin{align}
 &\Delta (k_{\rm peak}/\tilde{\beta}) \simeq 0.029 \,,
 \\
 &\frac{k_{\rm peak}}{\tilde{\beta}} \simeq 1.45 \,.
\end{align}
for $n=4$. 

These results reinforce the universality of the GW spectrum: its shape and peak position are largely insensitive to the specific nucleation history, depending primarily on the dimensionless ratio $k/\tilde{\beta}$.

\subsection{Case with power-exponential nucleation rate}
\label{sec:result_exp}

Finally, we consider a power-exponential nucleation rate: 
\begin{align}
 \tilde{\Gamma}(\tau) 
 &= \tilde{\Gamma}_0 \lmk \tilde{\beta}' \tau \rmk^n e^{\tilde{\beta}' \tau} \,,
 \label{eq:nucleation-rate-exp}
\end{align}
with constants $\tilde{\Gamma}_0$, $\tilde{\beta}'$, and $n$~\cite{Linde:1981zj,Kamionkowski:1993fg}. The parameter $\tilde{\beta}'$ matches \eq{eq:beta} in the limit $\tilde{\beta}/\tilde{H}_*\gg1$. 
For large $\tilde{\Gamma}_0/\tilde{\beta}'^4$, we have $\tilde{\beta}/\tilde{H}_*=\tilde{\beta}\tau_*\approx 4 + n$, in which case the phase transition completes while the power-law dependence dominates. 
For a fixed $n$, the spectrum again depends only on $k/\tilde{\beta}$ and $\tilde{\beta}\tau_*$ (or 
equivalently $\tilde{\Gamma}_0/\tilde{\beta}'^4$).

\begin{figure}
 \centering
 \includegraphics[width=0.95\hsize]{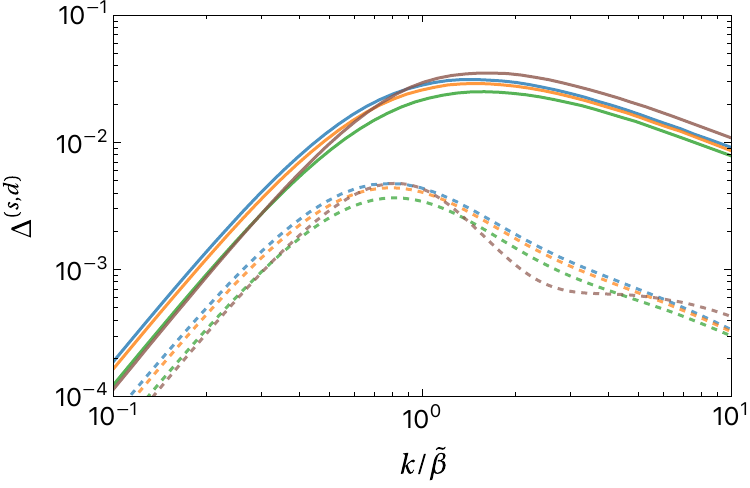}
 \\
 \vspace{0.3cm}
 \includegraphics[width=0.95\hsize]{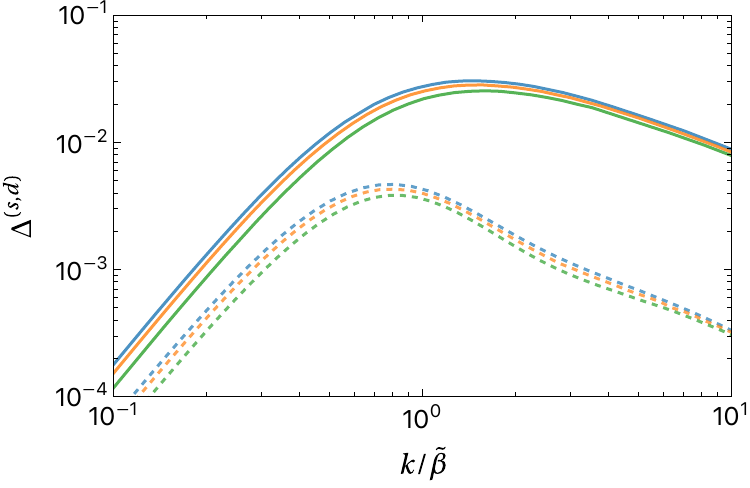}
 \\
 \vspace{0.3cm}
 \includegraphics[width=0.95\hsize]{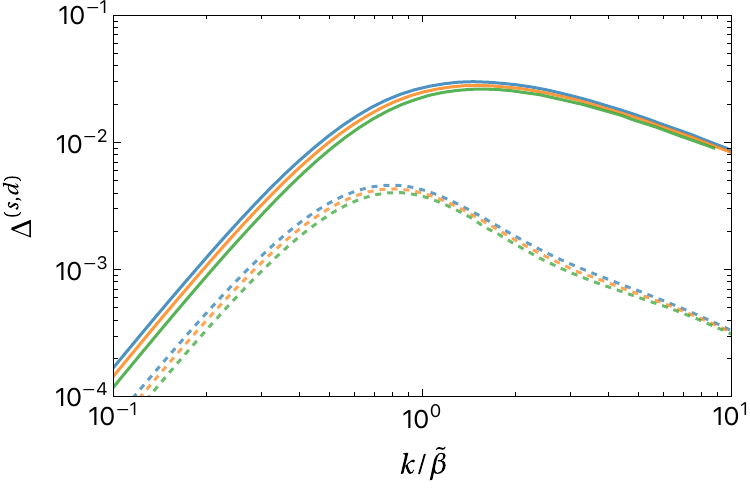}
 \caption{
 Same as Fig.~\ref{fig:delta1} but for a power-exponential nucleation rate. 
 We take $n=0$ (top), $2$ (middle), and $4$ (bottom)
 with $\tilde{\Gamma}_0/\tilde{\beta}'^4 = 10^{-15}$ (blue), $10^{-11}$ (orange), $10^{-7}$ (green), and $10^1$ (brown). 
 The case with $\tilde{\Gamma}_0/\tilde{\beta}'^4 =10^1$ is omitted in the lower two panels. 
}
 \label{fig:exp1}
\end{figure}

Figure~\ref{fig:exp1} shows $\Delta^{(s)}$ (solid) and $\Delta^{(d)}$ (dashed) spectra. In the top panel, we take $n = 0$ and choose $\tilde{\Gamma}_0/\tilde{\beta}'^4=10^{-15}$ (blue), $10^{-11}$ (orange), $10^{-7}$ (green), and $10^{1}$ (brown), corresponding to $\tilde{\beta}\tau_*\simeq 31$, $22$, $13$, and $4.1$, respectively.
In the middle and bottom panels, 
we take $n=2$ and $n=4$ with 
$\tilde{\Gamma}_0/\tilde{\beta}'^4=10^{-15}$ (blue), $10^{-11}$ (orange), and $10^{-7}$ (green). 
The case with $\tilde{\Gamma}_0/\tilde{\beta}'^4=10^{1}$ is omitted in these panels because such a large value of $\tilde{\Gamma}_0/\tilde{\beta}'^4$ effectively corresponds to a power-law nucleation rate for $n = 2$ and $n=4$.

\begin{figure}
 \centering
 \includegraphics[width=0.95\hsize]{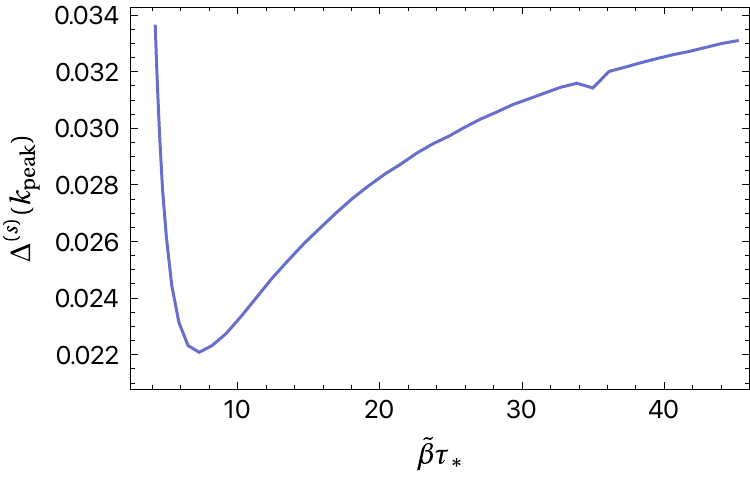}
 \\
 \vspace{0.3cm}
 \includegraphics[width=0.95\hsize]{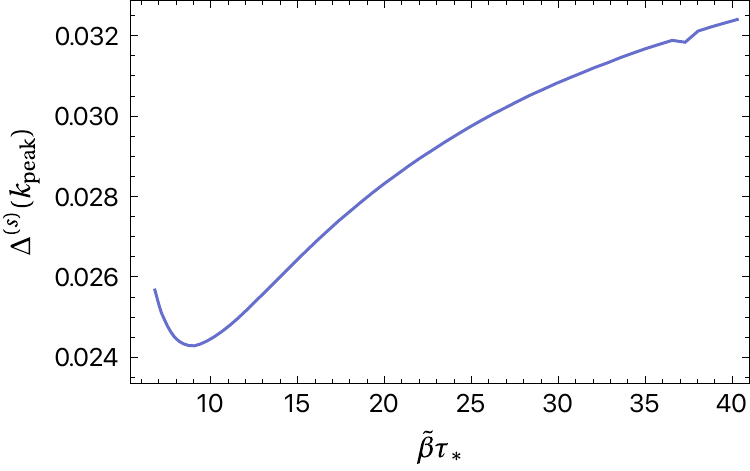}
 \\
 \vspace{0.3cm}
 \includegraphics[width=0.95\hsize]{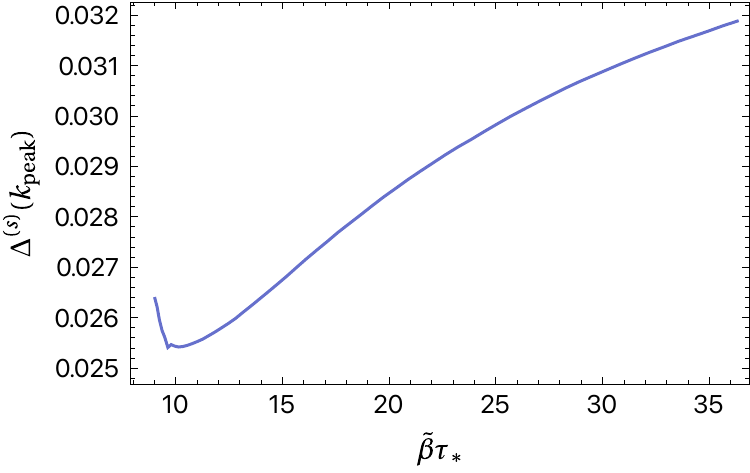}
 \caption{
 Same as the top panel of Fig.~\ref{fig:delta2} but for a power-exponential nucleation rate. 
 We take $n=0$ (top), $2$ (middle), and $4$ (bottom). 
}
 \label{fig:exp2}
\end{figure}

\begin{figure}
 \centering
 \includegraphics[width=0.95\hsize]{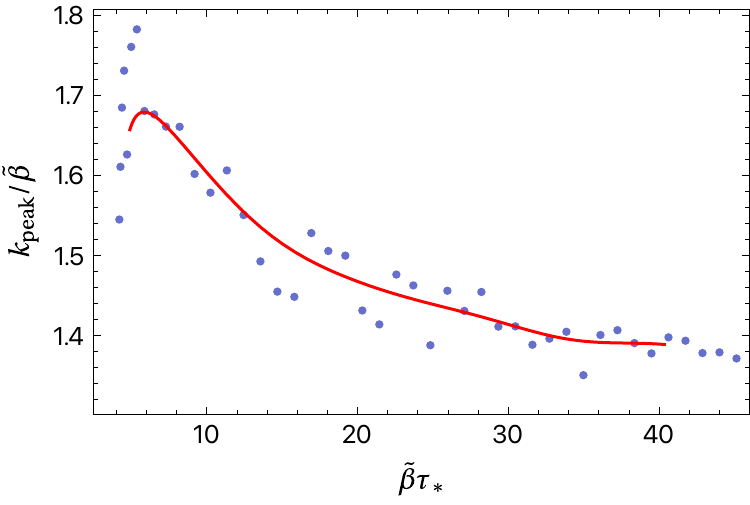}
 \\
 \vspace{0.3cm}
 \includegraphics[width=0.95\hsize]{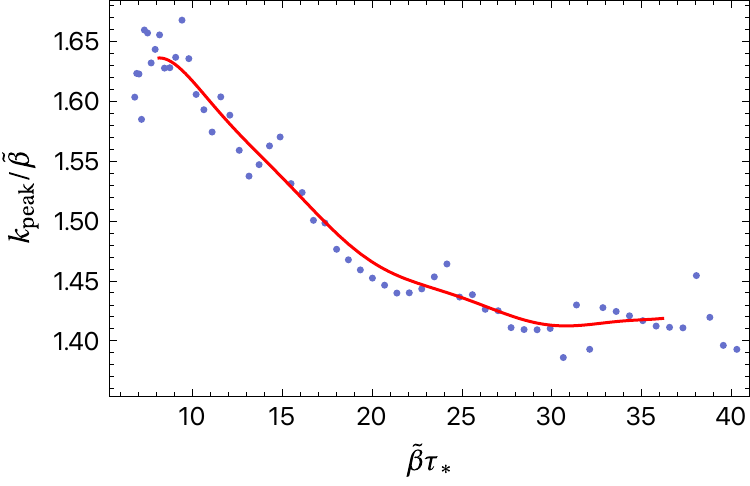}
 \\
 \vspace{0.3cm}
 \includegraphics[width=0.95\hsize]{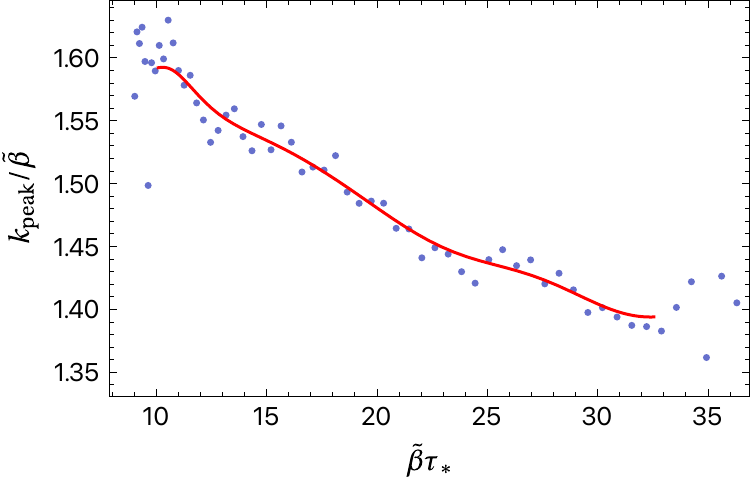}
 \caption{
 Same as the botoom panel of Fig.~\ref{fig:delta2} but for a power-exponential nucleation rate. 
 We take $n=0$ (top), $2$ (middle), and $4$ (bottom). 
}
 \label{fig:exp3}
\end{figure}

The peak amplitude $\Delta^{(s)}(k_{\rm peak},\tilde{\beta})$ and its position $k_{\rm peak}/\tilde{\beta}$ are plotted in Fig.~\ref{fig:exp2} and Fig.~\ref{fig:exp3} as functions of $\tilde{\beta}\tau_*$. 
Across all these cases, spectra plotted against $k/\tilde{\beta}$ remain remarkably similar up to an $\mathcal{O}(1)$ factor. Specifically,
\begin{align}
 &\Delta^{(s)}(k_{\rm peak}, \tilde{\beta}) \simeq 0.022\,\text{-}\,0.035 \,,
 \\
 &\frac{k_{\rm peak}}{\tilde{\beta}} \simeq 1.4\,\text{-}\,1.7 \,,
\end{align}
for $\tilde{\beta}\tau_* \gtrsim 4$ and $n=0$. 
The behavior changes when $\tilde{\beta} \tau_*$ becomes small, since the exponential term becomes subdominant in this regime, whereas it dominates for larger $\tilde{\beta} \tau_*$. 
Despite the distinct functional dependence, the peak amplitude and frequency vary only mildly with $\tilde{\beta}\tau_*$, consistent with the above results.

We conclude that the FLRW-spectrum can be reliably estimated from Minkowski-spacetime results by replacing $k$, $\beta$, and $H_*$ with their \emph{conformal} counterparts defined at the characteristic collision time $\tau_*$, up to an overall $\mathcal{O}(1)$ factor. 
This provides a simple and robust criteria for estimating GW signals from supercooled phase transitions in an expanding Universe.

\begin{figure}
 \centering
 \includegraphics[width=0.95\hsize]{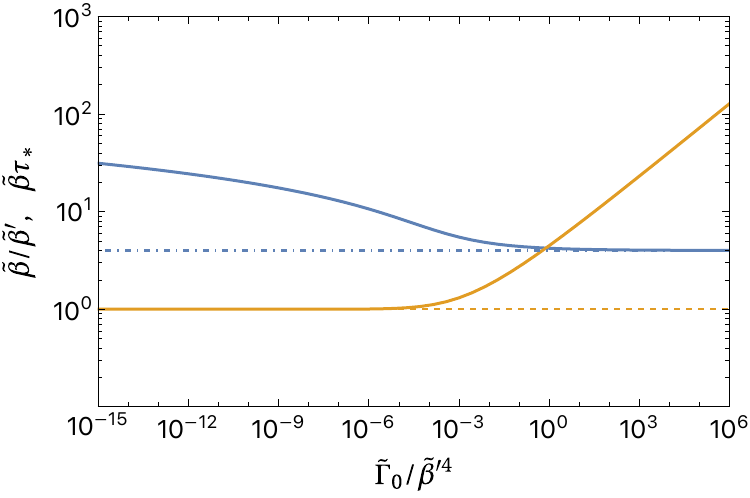}
 \caption{
 $\tilde{\beta}/\tilde{\beta}'$ (orange curve) and $\tilde{\beta} \tau_*$ (blue curve) as functions of $\tilde{\Gamma}_0 / \tilde{\beta}'^4$ for the case of $n=0$. 
 The dashed and dot-dashed lines indicate the asymptotic values.
}
 \label{fig:betatauf}
\end{figure}

Finally, we comment on how the GW spectrum can be overestimated by a naive extrapolation from the Minkowski-spacetime limit in the case $\beta/H_* = \tilde{\beta}/\tilde{H}_* = \mathcal{O}(1)$, using the example with $n=0$.
The key issue lies in the difference between our definition of $\tilde{\beta}$ and the parameter $\tilde{\beta}'$ appearing in the exponent of the bubble nucleation rate.
One might naively estimate the inverse (conformal) duration of the phase transition as $\tilde{\beta}'^{-1}$ (or $\beta'^{-1}$ for the physical duration). However, this identification is not appropriate when $\tilde{\beta}/\tilde{H}_* = \mathcal{O}(1)$, because the effects of cosmic expansion make the difference between $\tilde{\beta}$ and $\tilde{\beta}'$ significant.
The discrepancy, together with the value of $\tilde{\beta}/\tilde{H}_*$ ($=\tilde{\beta}\tau_*$), is shown in Fig.~\ref{fig:betatauf} as the orange and blue curves, respectively, plotted as functions of $\tilde{\Gamma}_0/\tilde{\beta}'^4$.
A small value of $\tilde{\Gamma}_0/\tilde{\beta}'^4$ corresponds to a large $\tilde{\beta}/\tilde{H}_*$, in which case $\tilde{\beta} \simeq \tilde{\beta}'$.
In contrast, in the parameter region where $\tilde{\beta}/\tilde{H}_* = \mathcal{O}(1)$ (or equivalently for relatively large $\tilde{\Gamma}_0/\tilde{\beta}'^4$), $\tilde{\beta}'$ significantly underestimates $\tilde{\beta}$.
As a consequence, the GW amplitude is overestimated if it is assumed to scale as $\propto \tilde{\beta}'^{-2}$, whereas the correct scaling is $\propto \tilde{\beta}^{-2}$ (see \eq{eq:Omega_Delta}).

\section{Discussion and conclusion}
\label{sec:discussion}

We present an analytic formulation of bubble collisions for runaway walls under the thin-wall and envelope approximations, taking into account the effects of cosmic expansion while neglecting the backreaction on the background metric.
We find that the spectral shape, amplitude, and peak frequency of the GW signal are largely insensitive to details of the bubble nucleation history once the parameter dependence is factorized as
$\Omega_{\rm GW}\propto(\tilde{\beta}/\tilde{H}_*)^{-2}(\kappa_*\alpha_*)^2$, where the subscript $*$ denotes quantities evaluated at the end of the phase transition. In the radiation-dominated scenarios considered here, the ratio satisfies $\tilde{\beta}/\tilde{H}_*\!\ge\!3$, implying an upper bound on the GW amplitude even for strongly supercooled transitions. Using $\Delta=\mathcal{O}(10^{-2}\text{--}10^{-1})$ from our results, we obtain a present-day peak bound
$\Omega_{\rm GW} h^2 \lesssim 10^{-7}\,\kappa_*^2\alpha_*^2$.

The physical origin is simple. In a strongly supercooled transition the characteristic bubble size at percolation is $\mathcal{O}(H_*^{-1})$ irrespective of the nucleation profile, fixing both the time/length scales sourcing tensor modes and the available energy density $\sim \kappa_*\alpha_*\rho_{\rm tot}\propto \kappa_*\alpha_* H_*^2$. Consequently, the GW amplitude scales solely as a power of $H_*$ (modulo the trivial factor $(\kappa_*\alpha_*)^2$ and $G$), and the spectrum peaks at $k\sim H_*$, i.e., $k/\tilde{\beta}=\mathcal{O}(1)$ when $\tilde{\beta}/\tilde{H}_*=\mathcal{O}(1)$.

Our analytic expressions also permit controlled expansions, such as large/small $k$ and the Minkowski limit, by expanding the integrands in the relevant small parameters. These analyses, together with the full derivation for the full analytic formula, are presented in a companion paper~\cite{Yamada:2025cfr}.

\

\section*{Acknowledgments}
We thank Ryusuke Jinno for fruitful discussions throughout this work. 
This work is supported by JSPS KAKENHI Grant Number 23K13092.


\bibliography{ref}


\end{document}